\documentclass[pra,twocolumn, showpacs,superscriptaddress]{revtex4}

\usepackage{amssymb}
\usepackage[dvips]{graphicx}
\usepackage{color}
\usepackage{amsmath}
\usepackage{amsthm}

\newcommand{\bra}[1]    {\langle #1|}
\newcommand{\ket}[1]    {| #1 \rangle}

\newtheorem{thm}{Theorem}[section]

\newtheorem{lem}[thm]{Lemma}

\begin{document}

\title{Singlet Generation in Mixed State Quantum Networks}

\author{S.~Broadfoot}
\affiliation{Clarendon Laboratory, University of Oxford, Parks Road, Oxford OX1 3PU, United Kingdom}
\author{U.~Dorner}
\affiliation{Clarendon Laboratory, University of Oxford, Parks Road, Oxford OX1 3PU, United Kingdom}
\author{D.~Jaksch}
\affiliation{Clarendon Laboratory, University of Oxford, Parks Road, Oxford OX1 3PU, United Kingdom}
\affiliation{Centre for Quantum Technologies, National University of Singapore, 117543, Singapore}

\date{\today}

\begin{abstract}
  We study the generation of singlets in quantum networks with nodes
  initially sharing a finite number of partially entangled bipartite
  mixed states. We prove that singlets between arbitrary nodes in such
  networks can be created if and only if the initial states connecting
  the nodes have a particular form. We then generalize the method of
  entanglement percolation, previously developed for pure states, to
  mixed states of this form. As part of this, we find and compare
  different distillation protocols necessary to convert groups of
  mixed states shared between neighboring nodes of the network into
  singlets. In addition, we discuss protocols that only rely
  on local rules for the efficient connection of two remote
  nodes in the network via entanglement swapping. Further improvements
  of the success probability of singlet generation are developed by
  using particular forms of `quantum preprocessing' on the network.
  This includes generalized forms of entanglement swapping and we show
  how such strategies can be embedded in regular and hierarchical
 quantum networks.
\end{abstract}

\maketitle

\section{Introduction \label{sec1}}

Quantum entanglement is one of the most notable features of quantum
systems and has been accepted as a key resource for quantum
information processing~\cite{N&C}. The distribution of entanglement
through quantum networks is therefore essential for the future of a
variety of applications, ranging from quantum cryptography to quantum
teleportation and distributed quantum
computing~\cite{PhysRevLett.70.1895}. However, the generation of these
entangled states faces a severe obstacle. Quantum channels such as
free-space transmission or optical fibers are prone to loss and
decoherence. This causes the desired maximally entangled states to
degrade into mixtures and limits the distance over which the quantum
information can be sent directly.  To overcome these problems `quantum
repeater' schemes have been
proposed~\cite{Briegel98,Duan01,Duer99,Childress05,Hartmann07,Dorner08}
which make use of the ability to `purify'~\cite{Bennett96a,Deutsch96}
and `swap'~\cite{Zukowski93,Bose99} entanglement to maintain a high
fidelity throughout. Quantum repeaters are a promising tool for
entanglement distribution, particularly since the amount of required
physical resources increases only polynomially with the
distance~\cite{Duer99}, but operate in a 1D setup of network nodes.
Real networks are typically two-(or higher) dimensional and it is
therefore desirable to study if entanglement distribution can be made
more efficient in these cases.

A scheme for entanglement distribution in higher dimensional networks
was recently proposed by Ac{\'i}n et al.~\cite{Acin07} in which ideas
from classical bond percolation have been applied to regular, i.e.
lattice-shaped, quantum networks. The scheme makes use of the
networks' connectivity and allows for the generation of maximally
entangled singlet states between arbitrary points of the network, with
a probability that is independent of their separation.  The only
requirement is that the nodes are initially connected by bipartite
{\em pure} states with sufficiently high entanglement. The restriction
to pure states was made since a pure, partially entangled state can be
converted into a singlet with finite probability via local operations
and classical communication (LOCC)~\cite{Vidal99} which is essential
for the bond percolation protocol: Initially one attempts to convert
all bipartite pure states into a singlet which, in each case, succeeds
with a certain probability. If this singlet conversion probability
(SCP) exceeds a lattice-geometry-dependent threshold, arbitrarily
large clusters of singlet-connected nodes form which can successively
be connected via entanglement swapping. In this way we can create a
singlet between arbitrarily remote nodes in the network. However, it
was pointed out in~\cite{Acin07} that this process, known as Classical
Entanglement Percolation (CEP), is not optimal since certain quantum
preprocessing schemes applied to the network can improve the
SCP~\cite{Acin07,Perseguers08,Lapeyre09,Eisert07,cuquet09,perseguers09b},
and thus it is possible to apply bond percolation to lattices in which
this would otherwise not be possible.

Clearly, the assumption of having a pure-state network is an idealization and
in any practical situation the states connecting the nodes of the
network will be mixed. In~\cite{Broadfoot09} the idea of entanglement
percolation was applied to mixed states for the first time. In this
paper we elaborate and extend the ideas presented
in~\cite{Broadfoot09}. The networks we consider are composed of nodes,
each of which can consist of several qubits, and may be connected by a
finite number of bipartite mixed states (see Fig.~\ref{fig1}).
We aim to create a {\em perfect} singlet between two arbitrary nodes
in the network using a {\em finite} amount of resources, i.e. a finite
number of initial states which are converted into a singlet, which
distinguishes our and other entanglement percolation schemes from,
e.g., the quantum repeater protocol where one aims to generate a state
with high but non-unit entanglement fidelity. Particularly we
structure the paper as follows.

In Sec.~\ref{sec2} we prove a necessary and sufficient condition that
a perfect singlet can be generated in a network of arbitrary geometry
the nodes of which are initially connected by bipartite mixed qubit
states. We show that singlet generation between two nodes is possible
{\em if and only if} they are connected by at least two `paths'
consisting of a particular class of states. These states arise
naturally in systems undergoing amplitude damping. Thus our result is
not only of theoretical but also of practical relevance.
Unfortunately, the proof does not deliver an efficient scheme for
singlet generation. We therefore specialize in the remaining sections
on networks with regular geometry, i.e. lattices in 2D and 3D and
devise generalizations of entanglement percolation to the mixed states
described in Sec.~\ref{sec2}.

In Sec.~\ref{sec3} we briefly summarize the idea of classical
entanglement percolation with pure states.

In Sec.~\ref{sec4} we extend the concept of classical entanglement
percolation to mixed states. To this end we consider regular networks
where each node is connected to its neighboring nodes by a finite
number of the mixed states introduced in Sec.~\ref{sec2}. We present
two different distillation protocols which are used to convert these
states into a singlet with a probability above the percolation
threshold of a variety of lattice geometries. After the distillation,
clusters of singlet-connected nodes emerge and we aim to create a
singlet between two nodes in such a cluster by successive application
of entanglement swapping. By communicating classically each node can
determine if singlets exist between it and its neighboring nodes. This
information can be communicated and stored classically in a central
data processor. Typically one would then use this information to apply
a path-finding algorithm which locates a suitable `path' of singlets
before swapping operations are performed. As an alternative to this we
discuss a classical and a quantum protocol which merely require
classical communication between neighboring nodes and basic computing
within each node.  The quantum protocol relies on the formation of
many-qubit GHZ states via local operations and classical
communication with neighboring nodes and subsequent measurements at
all nodes except the ones to be left in the final singlet.

In Sec.~\ref{sec5} we show that the idea of `quantum preprocessing' as
it was successfully applied in pure state networks can be generalized
to mixed states. In particular we devise a number of strategies on
small networks which improve the SCP, and we show that these smaller
networks can be embedded into larger networks to enable CEP which
would otherwise not be possible. Furthermore, we discuss `hierarchical
schemes', i.e. networks which are defined iteratively and were first
discussed in~\cite{Perseguers08,Lapeyre09}. Also in these cases it
turns out that quantum methods outperform classical percolation.
Finally, in Sec.~\ref{sec6} we summarize and conclude.

\section{Singlet Generation within an Arbitrary Mixed State Network \label{sec2}}
\begin{figure}[t]
  \centering\includegraphics[width=6cm]{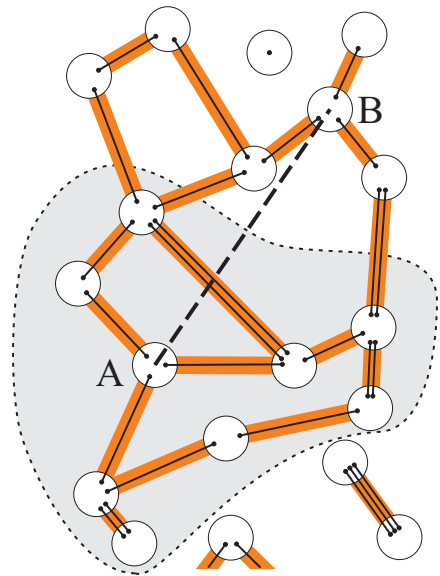}
\caption{%
  Mixed-state quantum network. Qubits in a node (circles) may be
  connected by bonds (thick lines), i.e. they share mixed entangled
  states, `edges', (solid, black lines) of qubits (black dots) with
  other nodes. When two `paths' of states of the
  form~(\ref{eq:state1}) connect $A$ and $B$ a singlet (dashed line)
  can be created with finite probability. This is proven by
  partitioning the nodes into two groups with one containing A (shaded
  region) and the other B.  For it to be possible to generate a
  singlet between A and B these groups must be linked by at least two
  states of the form~(\ref{eq:state1}) for all possible partitions.}
\label{fig1}
\end{figure}

In this section we consider quantum networks of arbitrary geometry as
shown in Fig.~\ref{fig1} where the qubits in the nodes are `connected'
by bipartite mixed states to qubits in other nodes. We will call a
single bipartite mixed state an {\em edge} and the set of edges
directly connecting two nodes a {\em bond}. Note that an edge connects
exactly two qubits in different nodes. In the following we will prove
that the generation of a perfect singlet between two arbitrary nodes
$A$ and $B$ with finite probability in such a network is possible if
and only if there are at least two paths of states linking $A$ and $B$
which have, up to local unitaries, the form
\begin{equation}
\rho(\alpha,\gamma,\lambda) = \lambda \ket{\alpha,\gamma}\bra{\alpha,\gamma} + (1-\lambda)\ket{01}\bra{01},
\label{eq:state1}
\end{equation}
where $\ket{\alpha,\gamma} = \sqrt{\alpha} \ket{00} +
\sqrt{1-\alpha-\gamma}\ket{11} + \sqrt{\gamma}\ket{01}$ and
$0\le\lambda\le1$. We show this by separately proving a necessary and
sufficient condition which, together, prove the above statement.

{\em Necessary condition.} We split the network into two groups of
nodes, $\mathcal{A}$, containing A and a finite number of other nodes,
and $\mathcal{B}$, which consists of the rest of the network and
particularly contains B. These groups are linked by a finite number of
edges. A singlet can be
established with finite probability, via local operations in the
groups and classical communication between them, if and only if at
least two of the states have the form~(\ref{eq:state1}).
Appendix~\ref{app1} contains a concise proof of this fact based on~\cite{Jane02} which agrees
with the result of Ref.~\cite{Kent98}, that, in general, a singlet can
not be generated with a finite probability from a finite number of
mixed states.

With two states of the form~(\ref{eq:state1}),
$\rho(\alpha,\gamma,\lambda)$ and $\rho(\beta,\delta,\nu)$, we obtain
a singlet with a finite probability by first performing two C-NOT
gates locally, with the $\rho(\beta,\delta,\nu)$ state's qubits acting
as the target qubits. These target qubits are then measured in the
computational basis. If we find both qubits to be in the state
$\ket{1}$ we have generated a pure entangled state between the qubits
that originally corresponded to the $\rho(\alpha,\gamma,\lambda)$
state. We will refer to this measurement as the pure state conversion
measurement (PCM). The state formed is
\begin{equation}
\ket{\alpha'} \equiv \ket{\alpha',\gamma=0} = \sqrt{\alpha'} \ket{00} + \sqrt{1-\alpha'}\ket{11},
\label{eq:Simple Pre State}
\end{equation}
i.e. $\alpha'$ is a Schmidt-coefficient that has the value
\begin{equation}
\alpha'=\frac{\min (\alpha (1-\beta-\delta),\beta (1-\alpha-\gamma))}{\alpha (1-\beta-\delta)+\beta (1-\alpha-\gamma)}.
\label{eq:PCM-S}
\end{equation}
The probability that the PCM succeeds in generating this state is
given by
\begin{equation}
p_c = \lambda\nu (\alpha (1-\beta-\delta)+\beta (1-\alpha-\gamma)).
\label{eq:PCM-P}
\end{equation}
For identical states, i.e. $\alpha=\beta,\,\gamma=\delta$, the PCM
already yields a singlet. Otherwise the state can be transformed into
a singlet via the `Procrustean method'~\cite{Bennett96} that converts
any pure 2-qubit state $\ket{\alpha'}$ into a singlet $\ket{1/2}$ with
a probability $2\min(1-\alpha',\alpha')$. The total success
probability of generating a singlet is then given by the SCP
\begin{equation}
p_{conv} = 2\lambda\nu \min[ \alpha(1-\beta-\delta),\beta(1-\alpha-\gamma) ]
\label{eq:prob}
\end{equation}
which coincides with the optimal probability for creating a singlet
from two of these states~\cite{Jane02}.

We can perform this partition of the network in an arbitrary way, as
long as one group contains $A$ and the other contains $B$. To be able
to create a singlet between $A$ and $B$ via LOCC we must have at least
two states of the form (\ref{eq:state1}) in {\em all} possible
partitions. This gives us a necessary condition that to create a
singlet between two nodes with a non-zero probability there have to be
at least two distinct `paths' of edges of the form~(\ref{eq:state1})
connecting the corresponding nodes. In Fig.~\ref{fig1}(a) this is
indicated by two spatially distinct paths of bonds. The states of the
qubits that are not contained in this path are irrelevant and can
therefore be in arbitrary states.

\begin{figure}[t]
  \centering\includegraphics[width=5cm]{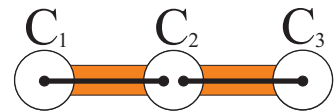}
\caption{%
  Basic arrangement for entanglement
  swapping.  Entanglement swapping involves a measurement in the Bell
  basis at node $C_2$ and classical communication between the nodes followed by local unitaries
  which causes $C_1$ and $C_3$ to become entangled.}
\label{fig2}
\end{figure}
{\em Sufficient condition.} In order to show this we make use of
entanglement swapping.  This operation can be performed in the setup
shown in Fig.~\ref{fig2} and consists of performing a measurement in
the standard Bell basis on the qubits located at $C_2$ and LOCC
which causes $C_1$ and
$C_3$ to become entangled. If the edges are of the
form~(\ref{eq:state1}), $\rho(\alpha,\gamma,\lambda)$ and
$\rho(\beta,\delta,\nu)$, then there are four possible outcomes. The
probabilities to obtain measurement outcomes corresponding to the Bell
states $\ket{\Psi^{\pm}}=(\ket{00}\pm\ket{11})/\sqrt{2}$ and
$\ket{\Phi^{\pm}}=(\ket{01}\pm\ket{10})/\sqrt{2}$ are
\begin{equation}
 p(\Psi_{\pm})=\frac{1}{2} (h_{\pm}\lambda \nu +(1-\beta -\delta )
(1-\lambda ) \nu +\alpha \lambda (1-\nu ))
\label{Equ:PSwapPMS1}
\end{equation}
and
\begin{equation}
 p(\Phi_{\pm})=\frac{1}{2}(g_{\pm}\lambda \nu +(1-\nu )
(1-\alpha \lambda)+(\beta +\delta ) (1-\lambda ) \nu),
\label{Equ:PSwapPMS2}
\end{equation}
where
\begin{align}
h_{\pm}=&\alpha \beta +(1-\alpha -\gamma)(1-\beta -\delta ) \nonumber \\
&+(\sqrt{\alpha  \delta}\pm \sqrt{\gamma (1-\beta -\delta )})^2, \\
g_{\pm}=&\gamma\beta +(1-\alpha -\gamma )\delta +(1-\alpha -\gamma )\beta \nonumber \\
&+(\sqrt{\gamma  \delta}\pm \sqrt{\alpha (1-\beta -\delta )})^2.
\end{align}
If we measure the qubits at $B$ to be in the states $\ket{\Psi^{\pm}}$
then we actually form another state,
\begin{equation}
 \rho \left(\frac{\alpha \beta}{h_{\pm}},\frac{(\sqrt{\alpha  \delta}\pm \sqrt{\gamma (1-\beta -\delta )})^2}{h_{\pm}},\frac{\lambda \nu h_{\pm}}{2p(\Psi_{\pm})}\right),
\label{Equ:SucSpecialSwapping}
\end{equation}
of the form~(\ref{eq:state1}) between $C_1$ and $C_3$. Unfortunately
for the other outcomes the states' form is not generally maintained.
Note that if $\delta=\gamma=0$ we can discard these cases by replacing
the state with $\ket{01}$ leading to an operation that transforms
$\rho(\alpha,0,\lambda)\otimes\rho(\beta,0,\nu)$ into
\begin{equation}
 \rho \left(\frac{\alpha \beta}{h_\pm
},0,\lambda \nu h_{\pm}\right),
\label{Equ:GenSpecialSwapping}
\end{equation}
which will be useful in Sec.~{\ref{sec5A}}. We can therefore create a
state of the form~(\ref{eq:state1}) with non-zero probability between two nodes of the network,
e.g. $A$ and $B$ in Fig.~\ref{fig1}, given that these nodes are
connected by a path consisting of states of the same form. Two such
states, originating from two paths, can then be converted into a
singlet, using a PCM and the Procrustean procedure. Unfortunately,
this scheme leads to an exponential decrease of entanglement
fidelity~\cite{Duer99}, and thus success probability, with the number
of swapping operations.  Hence it is not an effective solution to the
problem of long-distance entanglement distribution. In Sec.~\ref{sec4}
we will therefore introduce effective protocols which can be applied
in regular network geometries and succeed in creating a singlet with a
probability independent of distance.

Note that when entanglement swapping is done with pure states all of
the outcomes can be used, and if these outcomes occur with
probabilities $p_m$ the pure state $\ket{\tilde{\alpha}}$ with
\begin{equation}
 \tilde{\alpha} = \frac{1}{2} \left( 1+\sqrt{1-\frac{\alpha \beta (1-\alpha)(1-\beta)}{p_m^2}} \right)
\label{Equ:PSwappingR}
\end{equation}
is recovered by using classical communication and local unitaries.

\section{Classical Entanglement Percolation with Pure States \label{sec3}}

\begin{figure}[t]
  \centering\includegraphics[width=8cm]{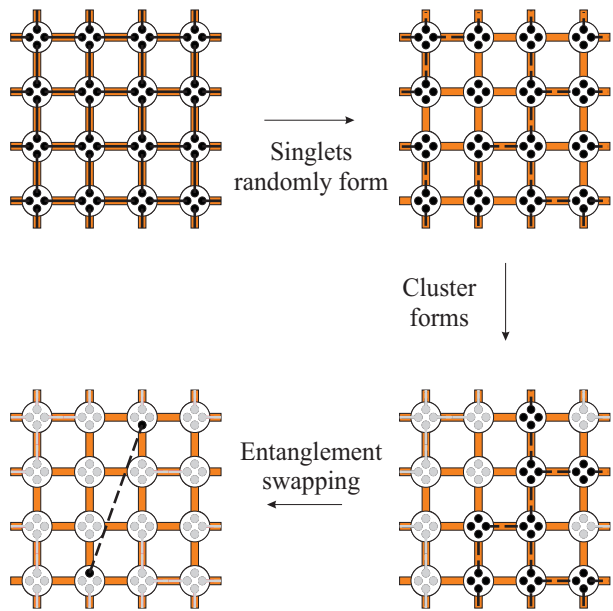}
\caption{%
  Illustration of classical entanglement percolation in a square
  network. Pairs of qubits (black dots) in neighboring nodes (circles)
  are in identical, pure, partially entangled states (solid, black
  lines). The percolation scheme involves these entangled states being
  converted into singlets (dashed lines) with probability $p$. If $p$
  exceeds the percolation threshold these form large clusters and we
  can obtain a singlet between any two qubits within a cluster by
  performing swapping operations. }
\label{fig3}
\end{figure}

In this section we will briefly review the use of percolation for
distributing singlets in pure state
networks~\cite{Acin07,Perseguers08}, known as classical entanglement
percolation (CEP). The procedure is based on classical bond
percolation, where we consider a regular lattice of nodes connected by
identical quantum states, as shown in Fig.~\ref{fig3}. A description
of classical bond percolation can be found in Ref.~\cite{book}. If the
nodes are connected by pure states of the form $\ket{\alpha}$
they can be converted into singlets using the Procrustean method with
a SCP $p=2\min(\alpha,1-\alpha)$. These singlets act as the bond in the bond
percolation model \footnote{Note that within the notation of this
  paper a `bond' corresponds to a set of states connecting two nodes
  and not to a perfect singlet} and are distributed randomly with a
probability $p$. The nodes that can be connected by a path of singlets
form a cluster. By using entanglement swapping (see Sec.~\ref{sec2})
we can then generate a singlet between any two nodes in the cluster.
In the theoretical case of an infinitely large lattice a cluster that
is infinite in extent forms if and only if $p>p_c$, where $p_c$ is a
lattice-dependent percolation threshold. This approximates the case
for large but finite lattices where the threshold becomes more
definitive as the size of the lattice increases. Values of $p_c$ for a
number of lattice geometries are given in Table~\ref{tab1}. If each
bond in a network consists of a single pure state $\ket{\alpha}$ we
can calculate a threshold for $\alpha$ given by $2\min(\alpha,1-\alpha)>p_c$. The
probability that a node belongs to the infinite cluster is known as
the percolation probability $\theta(p)$.  Two randomly chosen nodes
are both part of the infinite cluster with a probability $\theta(p)^2$
and thus can be connected over an arbitrary distance.
\begin{table}[ht]
\centering
\begin{tabular}{ll}
\hline
\hline
 Lattice & Threshold $p_c$ \\
 \hline
 2D Square & 0.5 \\
 2D Triangular & $2\sin(\pi/18) \approx 0.347$ \\
 2D Honeycomb & $1-2\sin(\pi/18) \approx 0.653$ \\
 3D Simple Cubic & $\approx 0.249$ \\
 3D Face-Centered Cubic & $\approx 0.120$ \\
 \hline
 \hline
 \end{tabular}
\caption{Threshold probabilities for various regular network geometries~\cite{book,Lorenz98}.
\label{tab1}}
\end{table}

It has been shown that CEP using pure states is not optimal and that
by performing particular quantum pre-processing steps, particularly
swapping operations on the lattice before converting to singlets,
improvements can be achieved. These improvements include obtaining a
geometry with a lower percolation threshold after the swapping
operation and splitting the lattice into two, so that a higher
percolation probability can be
obtained~\cite{Acin07,Perseguers08,Lapeyre09,Eisert07,cuquet09}.
Recently, another method, that transforms the initial bipartite network
into a probabilistic multipartite network, has also been shown to
yield an improvement~\cite{perseguers09b}.

\section{Classical Entanglement Percolation with mixed states \label{sec4}}

\begin{figure}[t]
  \centering\includegraphics[width=6cm]{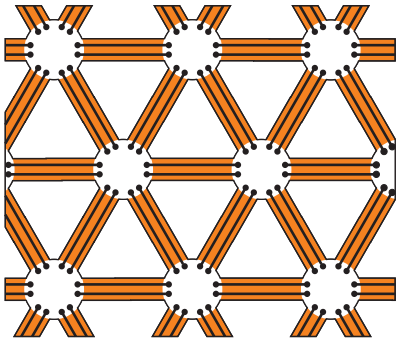}
\caption{%
  Triangular network. This is a simple 2D arrangement of
  PMSs in which CEP is possible.}
\label{fig4}
\end{figure}

In this section we extend CEP to mixed states. We consider regular
lattices, e.g.  triangular (see Fig.~\ref{fig4}), square, or even
lattices in higher dimensions. Bonds between network nodes are
composed of multiple edges which satisfies the necessary condition
proven in Sec.~\ref{sec2}. We assume that each bond is identical. When
these bonds contain at least two states of the form~(\ref{eq:state1})
they can be converted into singlets by PCM followed by the Procrustean
method. If the probability that a bond becomes a singlet exceeds the
percolation threshold CEP is achieved. In the remainder of the paper
we will assume that the states forming edges are of the
form~(\ref{eq:state1}) with $\gamma=0$.  Setting $\gamma=0$ is not a
major restriction but allows us to keep the equations manageable. All
protocols presented in this paper can also be performed if
$\gamma\ne0$. We will call states of the form~(\ref{eq:state1}) with
$\gamma=0$, i.e.
\begin{equation}
\rho(\alpha,\lambda)\equiv\rho(\alpha,\gamma=0,\lambda),
\end{equation}
purifiable mixed states (PMSs). Note that these states form the states
of two entangled atomic ensembles in the DLCZ quantum repeater
scheme~\cite{Duan01}.

\subsection{Distillation Procedures}

\subsubsection{Distillable Subspace Scheme}
\begin{figure}[t]
  \centering\includegraphics[width=3cm]{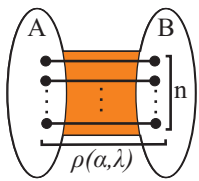}
\caption{%
  The purification setup consists of $n$ PMSs (solid lines) shared
  between two nodes $A$ and $B$. The aim is to distill these states into a singlet.}
\label{fig5}
\end{figure}

We assume that each pair of neighboring nodes is connected by $n$ PMSs
and our aim is to distill these into a singlet. The basic setup is
shown in Fig.~\ref{fig5}. To accomplish this we will use ideas
proposed in Ref.~\cite{Chen02}. Here the concept of a {\em distillable
  subspace} (DSS) is introduced as a subspace such that the local
projection of the system state into this space is pure and entangled.
Locating the DSS involves calculating the eigenvectors of the state
with non-zero eigenvalues. To simplify notation we will represent the
states at $A$ and $B$ using the decimal value of its binary form, i.e.
for example $\ket{00110}_A \ket{01001}_B = \ket{6}_A^d \ket{9}_B^d$.

As an example, in the case of $n=2$ identical states
$\rho(\alpha,\lambda)$ the eigenvalues and corresponding eigenvectors
are
\begin{eqnarray}
\lambda^2 :& \alpha \ket{0}_A^d \ket{0}_B^d  + \sqrt{\alpha(1-\alpha)}\ket{1}_A^d \ket{1}_B^d + \nonumber \\
&\sqrt{\alpha(1-\alpha)}\ket{2}_A^d \ket{2}_B^d  +  (1-\alpha) \ket{3}_A^d \ket{3}_B^d , \nonumber \\
\lambda(1-\lambda) :& \sqrt{\alpha} \ket{0}_A^d \ket{2}_B^d  + \sqrt{1-\alpha}\ket{1}_A^d \ket{3}_B^d ,\nonumber \\
\lambda(1-\lambda) :& \sqrt{\alpha} \ket{0}_A^d \ket{1}_B^d  + \sqrt{1-\alpha}\ket{2}_A^d \ket{3}_B^d ,\nonumber \\
(1-\lambda)^2 :& \ket{0}_A^d \ket{3}_B^d.  \label{Equ:GPEigenSys}
\end{eqnarray}
If this is acted on by the projective measurement $\ket{1}_A^d\bra{1}
+ \ket{2}_A^d\bra{2}$ at $A$ and $\ket{1}_B^d\bra{1} +
\ket{2}_B^d\bra{2}$ at $B$ the state remaining is
$(\ket{1}_A^d\ket{1}_B^d + \ket{2}_A^d\ket{2}_B^d)/\sqrt{2}$. Both of
these projective measurements only occur with probability
\begin{equation}
p_{n=2} = 2\lambda^2 \alpha(1-\alpha).
\label{eq:SCP2}
\end{equation}
Note that this is the same SCP as obtained for PCM [see
Eq.~(\ref{eq:PCM-S})]. For this example there is no choice between
entangled states to project out and if the original states are the
same a maximally entangled state is automatically obtained. For states
that are not identical this does not need to be the case.
\begin{figure}[t]
  \centering\includegraphics[width=8cm]{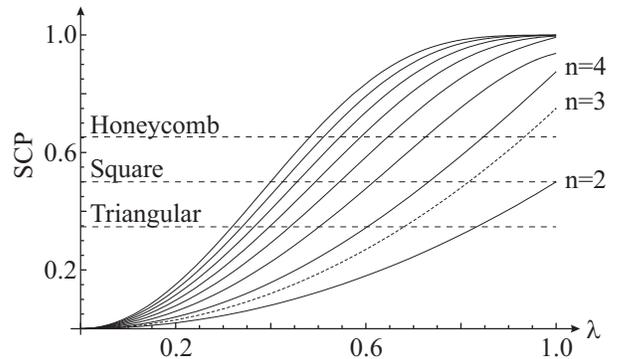}
\caption{%
  Singlet conversion probability for $n$-edged bonds using the
  recycling scheme for $\alpha = 1/2$ and $n=2,(3),4,6,8,10,12,14,16$
  (bottom to top). The $n=3$ line (dashed) corresponds to the DSS
  scheme.  The percolation thresholds for triangular (T), square (S)
  and honeycomb (H) lattices are given by the horizontal lines.  }
\label{fig6}
\end{figure}

An extension of this scheme to $n$ identical copies of PMSs
$\rho(\alpha,\lambda)$ yields the SCP
\begin{align}
&p_n = \sum_{m=0}^n  \lambda^{n-m} (1-\lambda)^m \binom{n}{m}   \nonumber \\
& \times\left( \sum_{k=1}^{n-m-1} \frac{\alpha^{n-m-k} (1-\alpha)^k \binom{n-m}{k} (\binom{n-m}{k}-1)}{\binom{n}{k}-1}\right).
\label{Equ:GP-ProbFn}
\end{align}
A derivation of this formula is given in Appendix~\ref{app2}. As a
particular example it is worthwhile to discuss the case of three
states in more detail. In this case the measurement at $A$ is given by a
Positive Operator Valued Measure (POVM) with the elements
\begin{align}
          &(\ket{0}_A^d\bra{0} + \ket{7}_A^d\bra{7}), \nonumber \\
          &(\ket{1}_A^d\bra{1} + \ket{2}_A^d\bra{2})/2, \nonumber \\
          &(\ket{1}_A^d\bra{1} + \ket{4}_A^d\bra{4})/2, \nonumber \\
          &(\ket{2}_A^d\bra{2} + \ket{4}_A^d\bra{4})/2, \nonumber \\
          &(\ket{3}_A^d\bra{3} + \ket{5}_A^d\bra{5})/2, \nonumber \\
          &(\ket{3}_A^d\bra{3} + \ket{6}_A^d\bra{6})/2, \nonumber \\
          &(\ket{5}_A^d\bra{5} + \ket{6}_A^d\bra{6})/2 .
\label{Eq:3GPMeasOps}
\end{align}
The measurement at $B$ then depends on this outcome and creates a
maximally entangled state with a certain probability.  The SCP is
obtained by setting $n=3$ in Eq.~(\ref{Equ:GP-ProbFn}) and is given by
\begin{equation}
p_{n=3} = 3 \lambda^2 \alpha (1-\alpha).
\label{Equ:GP-Three-Prob}
\end{equation}
Comparing this with the $n=2$ case [cf. Eq.~(\ref{eq:SCP2})] shows an
increase in the success probability which can be seen in
Fig.~\ref{fig6}, where the dashed line represents the SCP for
three identical states.

\subsubsection{Recycling scheme}\label{sec4A1}
The SCP using the DSS scheme does generally increase with increasing
$n$. However, the scheme does not make use of the available resources
in the best way. Indeed, the SCP $p_n$ can be
significantly improved by grouping $n$ identical PMSs into sets of $m$
and converting each of these sets into a singlet.
\begin{figure}[t]
  \centering\includegraphics[width=8cm]{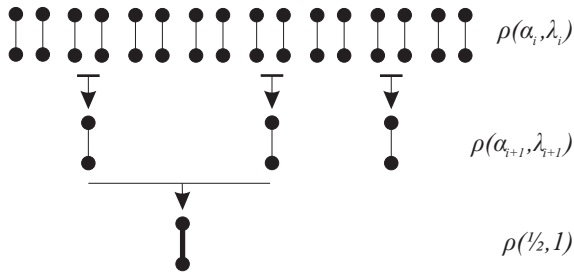}
\caption{%
  The recycling scheme consists of splitting the states into pairs
  that are then purified. If no singlets are successfully produced
  some of the states may still have been transformed into PMSs
  and given enough of these the process can be repeated.}
\label{fig7}
\end{figure}
For example for $m=2$ we apply the PCM as described in Sec.~\ref{sec2}
on pairs of states which converts them into singlets with a
probability given by Eq.~(\ref{eq:prob}). If this fails for a given
pair we may still find both measured qubits in the state $\ket{0}$ and have
generated another PMS. This PMS can then be used again in another
purification attempt. To be more precise, starting with $n$ copies of
a state $\rho(\alpha,\lambda)$ (with $\alpha\ge1/2$) we apply a
2-state purification protocol on groups of two. If no singlet is
obtained the procedure is repeated on the remaining PMSs as
illustrated in Fig.~\ref{fig7}.  The coefficients for the PMSs after
$k$ repetitions, when no singlet is created, are given by
\begin{eqnarray}
\alpha_k &=&  \frac{\alpha_{k-1}^2}{1-2\alpha_{k-1}+2\alpha_{k-1}^2}, \\
\lambda_k &=& \frac{\lambda_{k-1}^2(1-2\alpha_{k-1}+2\alpha_{k-1}^2)}{1-2\lambda_{k-1}+2\lambda_{k-1}^2(1-\alpha_{k-1}+\alpha_{k-1}^2)   }  ,
\end{eqnarray}
where $\alpha_0=\alpha$ and $\lambda_0=\lambda$.  For states of the
form $\rho(\alpha_k,\lambda_k)\otimes\rho(\alpha_k,\lambda_k)$ the
probability of obtaining a PMS is $c_k=1-2\lambda_k +
2(1-\alpha_k+\alpha_k^2)\lambda_k^2$. If the PCM yields two qubits
that are measured in different states the purification step between
the two PMSs has completely failed. The probability of this is given
by $f_k=2 \lambda_k (1-\lambda_k)$. The probability of not generating
a singlet using this {\em recycling protocol} on $n$ states of the
form $\rho(\alpha_i,\lambda_i)$ is then found to be
\begin{equation}
F_n(i) = \sum_{k=0}^{\lfloor \frac{n}{2} \rfloor} \left( \binom{\lfloor \frac{n}{2} \rfloor}{k} f_i^{\lfloor \frac{n}{2} \rfloor -k} c_i^k F_k(i+1) \right),
\label{Equ:Rec-iterativeFail}
\end{equation}
where $F_0(i) = 1$. Consequently, the probability of successfully
generating a singlet by applying the procedure to $n$ states of the
form $\rho(\alpha,\lambda)$ is $1-F_n(0)$ which is calculated
iteratively. Examples are shown in Figs.~\ref{fig6} for $\alpha=1/2$.
\begin{figure}[t]
  \centering\includegraphics[width=8cm]{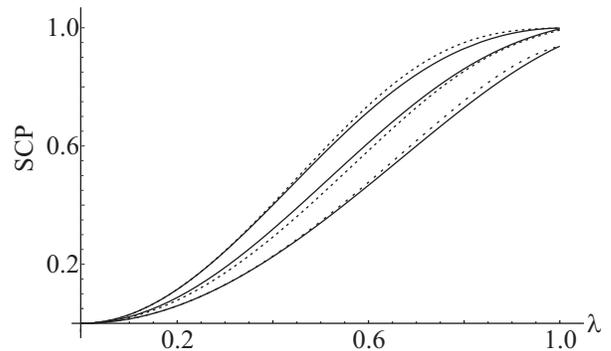}
\caption{%
  Success probabilities for the recycling schemes that split the
  states into pairs (dashed lines) and sets of three (solid lines).
  Shown are the SCPs for 6, 9 and 12 initial states (bottom to top)
  for $\alpha = 1/2$.}
\label{fig8}
\end{figure}

Obviously, the states do not necessarily need to be split into pairs.
For example we can separate all of the states into sets of three and
apply the three-state DSS distillation. In case of failure this can
yield a PMS state as well, which can then be used in later
distillation steps. There are a variety of ways to combine the
three-state distillation with the two-state recycling scheme. Here we
concentrate on the straightforward approach which only uses the
three-state distillation on every level of the recycling scheme. The
results are shown in Fig.~\ref{fig8}. As can be seen in most cases the
two-state recycling scheme has a higher chance of success and because
of this we will focus on the pairing arrangement in this paper.

%

\subsection{Percolation Thresholds}
Using the purification procedures described above we can apply CEP, as
described in Sec.~\ref{sec2}, for lattice networks with multi-edged
bonds. In most cases it is advantageous to use the two-state recycling
scheme, except for $n=3$ where the DSS scheme should be used. From
Fig.~\ref{fig6} it can be seen that the SCP increases with the number
of edges per bond and this allows for a larger range of values for
$\lambda$ and $\alpha$ such that CEP is successful. For double edged
bonds the optimal probability of generating a singlet is given by
\begin{equation}
0 \leq 2\lambda^2 \alpha(1-\alpha) \leq  1/2.
\label{Equ:2State-Prob}
\end{equation}
When the bonds are composed of three edges, i.e. three PMSs between nodes, we have
\begin{equation}
0 \leq 3\lambda^2 \alpha(1-\alpha) \leq  3/4.
\label{Equ:3State-Prob}
\end{equation}
By comparing these ranges to the percolation thresholds we see that
a basic successful setup is a double bonded triangular lattice
(see Fig.~\ref{fig4}). The double bonds can be converted to singlets
and if the chance of this is larger than the percolation threshold an
infinite cluster will form. A singlet can then be created between any
two nodes within the cluster. Thus percolation occurs if
\begin{equation}
2\lambda^2\alpha(1-\alpha) > 2\sin(\pi/18) \approx 0.347.
\end{equation}
However, the singlet conversion probability never exceeds $1/2$ for
two states. Therefore we would require more states in other
geometries. For example, if we have three-edged bonds between each
neighboring node we can apply CEP to a square lattice. This is because
there are parameters such that $3\lambda^2 \alpha(1-\alpha) > 1/2$.
Analogously, CEP is also possible in honeycomb lattices with three
edges per bond.
%
\subsection{Local Processing Strategies}
\begin{figure}[t]
  \centering\includegraphics[width=8cm]{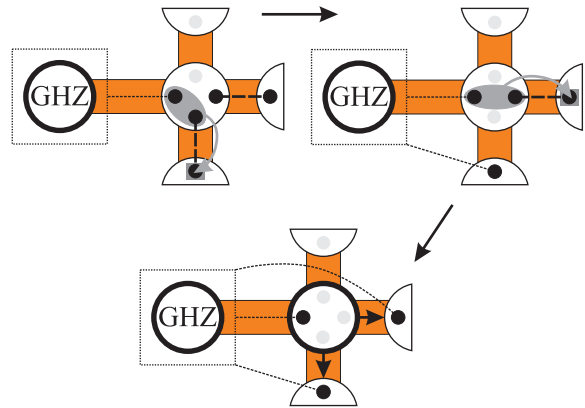}
\caption{%
  Procedure to join a node's singlets onto a GHZ state. Here the GHZ
  state is represented by a dotted box. Additional qubits (black dots) that are
  a part of the GHZ state are linked to the dotted box by a dotted line. A
  CNOT gate and measurement (both are represented by a shaded oval)
  are performed between the qubit already in the GHZ state and those
  that are part of a singlet (dashed lines). Each measurement outcome
  needs to be sent (gray arrow) to the other singlet qubit to
  perform a local unitary (shaded square). This extends the GHZ state
  to include qubits connected by edges to the node being attached.
  Once this has occurred for each qubit in a singlet the process is
  repeated by sending out a signal to repeat the step at each node
  that was linked by a singlet.}
\label{fig9}
\end{figure}

The process of creating singlets, randomly replacing the initial
network bonds, can be run if each node can only communicate
classically with their neighbors. Each node then knows if a qubit
that it contains is part of a singlet after this procedure has
finished. This information can be stored classically within a node but
after the bonds are distilled we are faced by the problem of finding a
set of singlets that connect our requested nodes, $A$~and~$B$.

If all of the singlet generation data is collected by a `controller'
then an efficient path finding algorithm can be applied to determine a
suitable `path' of singlets linking the nodes. An example of a
suitable algorithm would be a Dijkstra scheme~\cite{Dijkstra} such as
the A* path finding algorithm~\cite{Hart}. The path information can
then be used to instruct the correct nodes to perform swapping. The
swapping operations are performed in order from node $A$ to $B$, so
that the measurement outcomes only need to be communicated along the
chain, between neighboring nodes. However this procedure requires one
classical computer to have complete knowledge of the network. Instead,
it is interesting to note that this does not need to be the case as
there are algorithms which do not require any more classical
communication than this `controller' method, indeed they do not
require a central `controller' at all. This can be done not only classically
but also via a quantum algorithm using multipartite entanglement which
we will introduce below.

A classical path-finding method would use a type of breadth-first search algorithm called a burning
algorithm~\cite{Herrmann}. Node $A$ sends a `burning' signal to its
neighboring nodes connected by singlets. These nodes keep a record of
where they received the signal from and send out an identical signal
to the other nodes that they are connected to. We say that the node
has `burned'. If it has already received a signal from a different
node then the additional signal is ignored. This continues outwards
from $A$, `burning' the nodes. Once node $B$ receives the signal it
replies to the node it came from with a `swapping' message.  This node
can then perform a swapping operation and send another `swapping'
signal, together with the Bell-measurement outcome, back to the node
it received a `burning' signal from. The path can then be traced back
along the nodes with swapping performed at each step until node $A$ is
reached. Both $A$ and $B$ can determine if the protocol has been
successful. However, $A$ and $B$ may not be in the same cluster and
they do not know if the protocol has failed when the network is of
infinite size. This is not a problem for finite networks, containing
$N$ nodes, as $A$ and $B$ can time the steps taken and if these exceed
$2(N-1)$ they both know they are not in the same cluster.

Note that no extra information actually needs to be transmitted. We
can combine the burning algorithm with the process of transmitting the
distillation protocol information. For example, in a double edged
network of identical edges, $A$ can perform her PCM and if $\ket{1}$
is the outcome she assumes she has a singlet and sends a burning
signal to the node that would contain the singlet's other qubit. If a
node receives this signal it can perform its PCM and determine if
there is a singlet there. When there is and if it is the first
instance for the node it should record that entry qubit and repeat the
process, performing a PCM on the remaining qubits and sending signals
to those with the $\ket{1}$ outcome. Once $B$ receives a signal it can
check that a singlet has been created with a PCM and then send a
swapping signal back as before. During the swapping, a node can use
the Bell-measurement information received to indicate that a swapping
is required so no explicit `swapping' signal is required either. All
of this information transfer would have been necessary as well if a
controller algorithm would have been used. Hence the generation of the
singlet can be accomplished by defining rules for each node and
allowing them to run with nearest neighbor classical communication.
This is fundamentally different to the controller process and has made
use of parallel computation to find a path that no single node has
full knowledge of.

We will now consider an alternative, quantum algorithm that is based
on the burning algorithm and makes use of multipartite entanglement in
the network. The protocol starts after we attempted to convert all
bonds into singlets and every node has knowledge about its singlet
connections to nearest neighbors. We build up a progressively larger
multi-qubit GHZ state, defined by
$\ket{GHZ_n}=(\ket{0}_1...\ket{0}_n+\ket{1}_1...\ket{1}_n) /\sqrt{2}$,
spread between the `burned' nodes by adding qubits in each burning
step. Building up such a state requires joining two GHZ states,
$\ket{GHZ_n}$ and $\ket{GHZ_m}$, to create $\ket{GHZ_{n+m-1}}$ (note
that a singlet equals $\ket{GHZ_2}$). This is done by performing a
CNOT gate between a qubit in $\ket{GHZ_n}$ and a target qubit in
$\ket{GHZ_m}$, measuring the target qubit in the $Z$-basis,
communicating the measurement result to the other qubits in
$\ket{GHZ_m}$ and performing a unitary operation on them depending on
the outcome.
\begin{figure}[t]
  \centering\includegraphics[width=8cm]{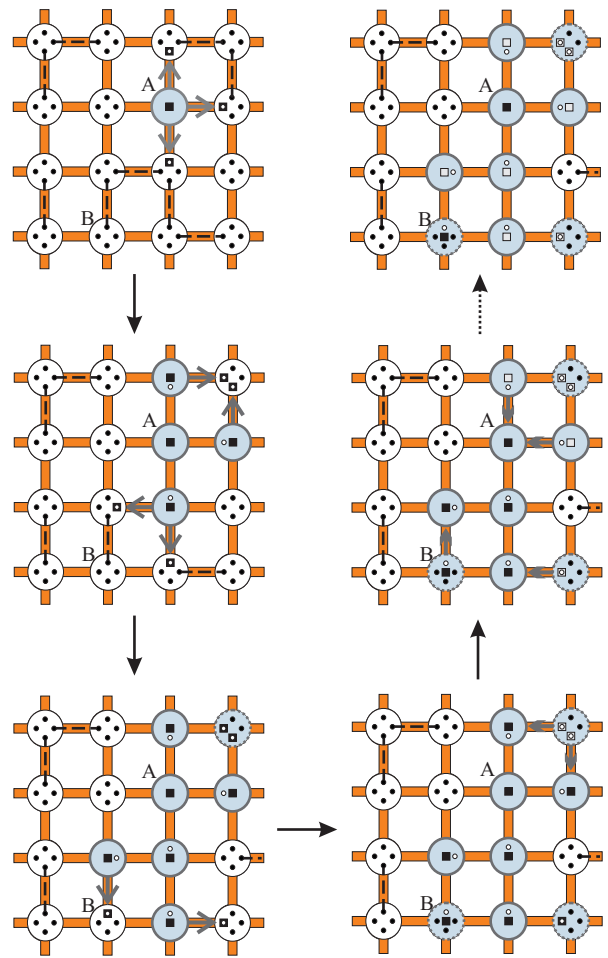}
\caption{%
  After the singlets are formed we can repeatedly extend a GHZ state
  from node $A$. This procedure uses the operation shown in
  Fig.~\ref{fig9} to add qubits to the GHZ state. The black squares
  depict qubits that are part of the GHZ state. Arrows represent a
  message to add nodes to the GHZ state along singlet paths. Each node
  keeps a record of the node from which it received this message from,
  symbolized here by a white dot. When a node cannot extend the GHZ
  state any further (highlighted by a dashed outline) it measures its
  qubits in the $X$ basis (open squares) and sends this information
  back towards $A$ (thin arrows) along the route recorded. Certain
  nodes are selected beforehand not to perform the measurement (here $A$ and $B$) and
  these will form the resulting GHZ state. At each node the incoming
  data can be combined and sent back along one path if the routes
  branch. Once this data returns to $A$ a phase operation can be
  performed on the qubit there to correct for any errors and the final
  GHZ state (here a singlet between $A$ and $B$) will remain.}
\label{fig10}
\end{figure}
Now we perform the same process as for the `burning algorithm',
however, as each node is `burned' it is connected to the GHZ state
spread over the previously burned nodes. The process to do this is
illustrated in Fig.~\ref{fig9} and consists of joining the singlets
partially contained in that node to the GHZ state. Within each node
one qubit is left entangled with the GHZ state. After this operation
has been run for a maximum of $N-1$ times all of the nodes in the
cluster containing $A$ have a qubit from a single GHZ state.

At each node a record is kept of the bond via which it has been
included into the GHZ state. If there is a singlet between two nodes
that are being burned then the singlet is ignored. Furthermore we add
the rule that whenever a node can not extend the GHZ state anymore $X$
basis measurements are performed along the recorded path back to $A$. This
removes a qubit from the GHZ state but introduces a phase error in the
remaining GHZ state depending on the outcomes of the measurement. The
information about these measurement outcomes has to be sent back along
the path to $A$. Whenever the route back branches, the measurement
outcome is sent in one way and a message corresponding to `no phase
error occurrence' is sent to the others. At each node the returning
process is paused until all of the bonds it sent a burning signal to
provide it with the phase information. At nodes $A$ and $B$ we do not
perform the $X$ measurement. Finally after $A$ receives all of the
phase information a phase correction can be performed and we obtain a singlet between $A$ and $B$. In
Fig.~\ref{fig10} an example is given to illustrate the protocol.

\section{Quantum Preprocessing \label{sec5}}

Despite being a very effective method, it is known that CEP in a
network of pure states can be improved by certain quantum
`pre-processing' strategies, and therefore CEP is not
optimal~\cite{Acin07,Perseguers08,Lapeyre09,Eisert07,cuquet09,perseguers09b}.
In the following we show that this is also the case in mixed-state
networks.

\subsection{Swapping procedure} \label{sec5A}

To start with we generalize the swapping arrangement shown in
Fig.~\ref{fig2} previously studied for pure
states~\cite{Perseguers08,Bose99}.  In this arrangement we have two
2-qubit states that both have a qubit in a common node. If the two
states are pure states $\ket{\alpha}$ and $\ket{\beta}$, with $\alpha\geq 1/2$ and $\beta\geq 1/2$, we can obtain
a singlet by swapping and then converting the resulting pure state
into a singlet with a total probability of $2
\min((1-\alpha),(1-\beta))$ which turns out to be the optimal
probability. Particularly CEP, which consists here of the Procrustean
method followed by entanglement swapping, always has a smaller SCP of
$4(1-\alpha)(1-\beta)$. Note that the optimal probability is equal to
that of converting the least entangled of the two bonds into a singlet
using the procrustean scheme~\cite{Bose99}.
\begin{figure}[t]
  \centering\includegraphics[width=8cm]{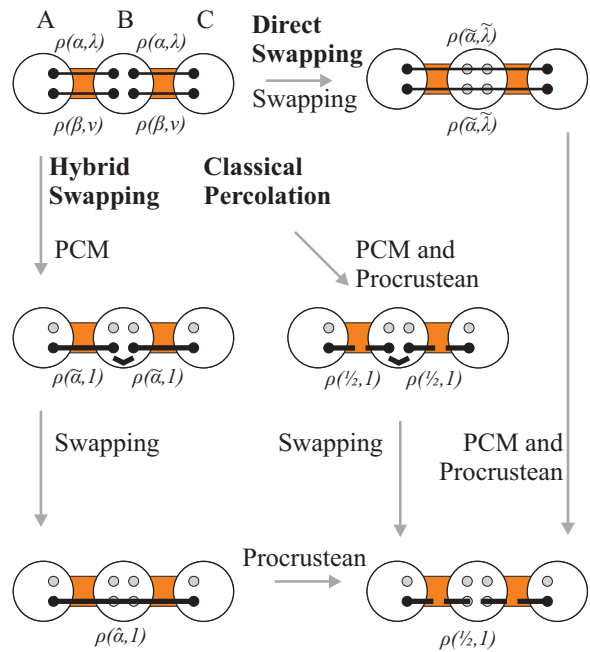}
\caption{%
  Three methods can be used to generate a singlet between two nodes
  $A$ and $C$ via an intermediate node $B$ in an arrangement with two
  edges per bond. The thick black lines indicate pure but not
  maximally entangled states.}
\label{fig11}
\end{figure}

To generalize this to mixed states we must consider double-edged bonds,
each consisting of two PMSs, as illustrated in Fig.~\ref{fig11}, since
otherwise singlet generation would not be possible. Introducing more
than one edge between the nodes allows us to concentrate the
entanglement at different stages which gives rise to three different
possibilities:
 \begin{description}
 \item[I] CEP - As previously described, the bonds are converted to
   singlets and then swapping is performed over the resulting states.
 \item[II] Direct swapping - This applies entanglement swapping twice
   and then the resulting states are converted into a singlet.
 \item[III] Hybrid swapping - Here we distill a state of higher
   entanglement in each bond (but not necessarily a singlet) leading,
   if successful, to a single (partially) entangled pure state in each
   bond. This is followed by entanglement swapping and the Procrustean
   scheme to create a singlet.
 \end{description}
 Each of these possibilities uses the swapping operation at different
 stages as illustrated in Fig.~\ref{fig11}. The exact implementations
 for the procedures depend on the types of states used. We will first
 apply each of them on a network of pure states and compare the SCPs.
 We then generalize to PMSs and show that direct and hybrid swapping
 can outperform CEP.

\subsubsection{Pure states}
If we start with bonds made of pure states $\ket{\alpha}$ and
$\ket{\beta}$ we must have a way to convert each bond into a singlet
in order to apply CEP(I). The method and highest possible probability
to accomplish this are given by Majorization~\cite{Nielsen01} with a
probability $p = \min (1,
2(1-\alpha\beta))$~\cite{Perseguers08,Lapeyre09}. CEP applies this
operation on each bond and if both bonds are converted into singlets
swapping can be performed and the operation is a success. Therefore
CEP succeeds with a probability $(\min(1,2(1-\alpha\beta)))^2$.

Our second method, direct swapping(II), is simply the application of
the procedure for bonds containing one edge twice. If either generates
a singlet the procedure succeeds. This gives a SCP of
$1-(1-2(1-\alpha))(1-2(1-\beta))$. There are adjustments we could
make, for example use the results of Majorization to convert both of
the states into a singlet with the highest possible probability, however
all of these have a smaller SCP than CEP for a range of parameters.

Finally, the hybrid swapping(III) method concentrates each bond to one
pure state, $\ket{\max (1/2, \alpha\beta)}$, with certainty. This
concentration procedure is also found using results from Majorization
theory~\cite{Nielsen01}. Afterwards there is one pure state in each
bond, as discussed previously, and we can then perform the strategy
with optimal success probability $\min(1,2(1-\alpha\beta))$, i.e.
swapping over the pure states followed by the Procrustean method. We
can actually consider the setup as a bipartite system between $A$ and
$BC$. The Majorization results then give the best possible probability
of generating a maximally entangled 2-qubit state between these
systems as $\min(1, 2(1-\alpha\beta))$ which means that it must be
the highest possible probability for any method to succeed.

\begin{figure}[t]
  \centering\includegraphics[width=8cm]{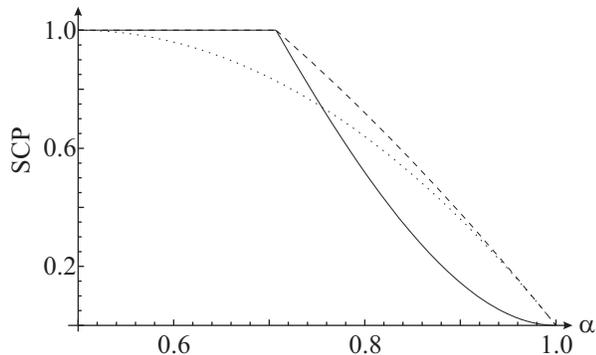}
\caption{%
  Comparison of the three methods described in the text for creating a
  singlet between nodes $A$ and $C$ in the setup shown in
  Fig.~\ref{fig11} for pure states. Shown are the success
  probabilities if the bonds are made up initially of the states
  $\ket{\alpha}$ and $\ket{1/2}$ for CEP (solid line), direct swapping
  (dotted line) and hybrid swapping (dashed line).}
\label{fig12}
\end{figure}

Figure~\ref{fig12} shows the probabilities in all three cases and we
can see that CEP is outperformed for a vast range of parameters by
both other strategies. In hybrid swapping (III), we have used
multi-edged bonds to create pure states with the highest probability
before applying entanglement swapping. We will refer to all strategies
that have this property as `hybrid'. This probability is unity for
initial pure states but for mixed states the initial conversion of
bonds to pure states is probabilistic, so when the conversion fails
the bond is destroyed.

\subsubsection{Purifiable Mixed States}

We will now investigate if similar improvements can be obtained with
PMSs, i.e. if the bonds between the nodes are composed of
$\rho(\alpha,\lambda)$ and $\rho(\beta,\nu)$. Again we will see that
hybrid swapping provides the highest SCP.

\begin{description}
\item[I] CEP

  The classical percolation scheme involves performing a PCM described
  in Sec.~\ref{sec2} followed by the procrustean protocol on both bonds
  and each succeeds with a probability given by Eq.~(\ref{eq:prob})
  which simplifies to
\begin{equation}
p_{conv}=2\lambda\nu\min (\alpha (1-\beta),\beta (1-\alpha)).
\label{Equ:MSwap-CPerc-SProb}
\end{equation}
To perform a swapping operation yielding a singlet, between nodes $A$
and $C$ we must succeed for both bonds which gives the total
chance of success
\begin{equation}
p_{CEP}=(2\lambda\nu\min (\alpha (1-\beta),\beta (1-\alpha)))^2,
\label{Equ:MSwap-CPerc-Prob}
\end{equation}
by simply squaring Eq.~(\ref{Equ:MSwap-CPerc-SProb}). In this case the
swapping operation is the final step of the protocol.

\item[II] Direct swapping

  In our 2-edged setup we perform the swapping operation introduced in
  Sec.~\ref{sec2} twice and there are two choices to do this if the
  states are not identical. Either we perform the swapping over the
  identical states $\rho(\alpha,\lambda)\otimes\rho(\alpha,\lambda)$
  or we perform the operation on the states
  $\rho(\alpha,\lambda)\otimes\rho(\beta,\nu)$.  When we swap over
  identical states we obtain the state
\begin{equation}
\rho\left(\frac{\alpha^2}{1-2\alpha+2\alpha^2},\lambda^2 (1-2\alpha+2\alpha^2 )\right),
\label{Equ:MSwap-NaiveIdstate}
\end{equation}
together with a further state where $\beta$ is replacing $\alpha$ and
$\nu$ is replacing $\lambda$. Note that
Eq.~(\ref{Equ:MSwap-NaiveIdstate}) is obtained by setting
$\gamma=\delta=0$ and $\alpha=\beta,\,\lambda=\nu$ in
Eq.~(\ref{Equ:GenSpecialSwapping}). This pair of states can then be
transformed into a singlet with a probability
\begin{equation}
p_{d^*}=2\lambda^2 \nu^2 \min(\alpha^2 (1-\beta)^2,\beta^2 (1-\alpha)^2)
\label{Equ:MSwap-NaiveProb}
\end{equation}
which is calculated using Eq.~(\ref{Equ:MSwap-CPerc-SProb}). In the case where we swap over
non-identical states $\rho(\alpha,\lambda)\otimes\rho(\beta,\nu)$ we obtain two
states of the form~(\ref{Equ:GenSpecialSwapping}) with $\gamma=\delta=0$. These can be
converted into a singlet with probability
\begin{equation}
p_d = 2 \lambda ^2 \nu ^2\alpha \beta (1-\alpha )(1-\beta ).
\label{Equ:MSwap-NaiveProb2}
\end{equation}
This is always larger than $p_{d^*}$ and thus swapping with non-identical states should be preferred.

\begin{figure}[t]
  \centering\includegraphics[width=8cm]{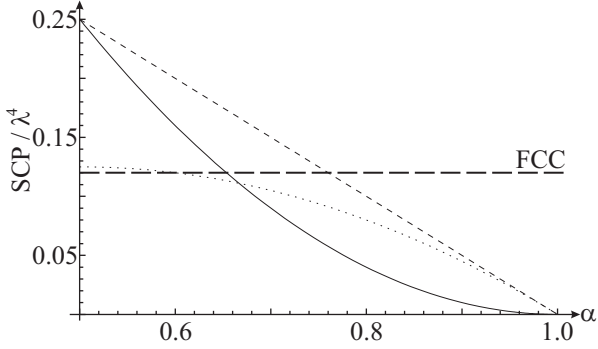}
\caption{%
  Success probability to generate a singlet between the end nodes
  of the swapping setup shown in Fig.~\ref{fig11} for the classical
  scheme (solid line), direct swapping (dotted line), and the
  hybrid scheme (dashed line). Each bond initially contains the states
  $\rho(\alpha,\lambda)$ and $\rho(1/2,\lambda)$. We have indicated
  the percolation threshold of a face-centered cubic network.}
\label{fig13}
\end{figure}

\item[III] Hybrid swapping

  The hybrid method requires a concentration procedure to be performed
  (yielding a single pure state in each bond) which is given here by
  PCM. However, in contrast to the pure state case discussed above, if
  $\alpha=\beta$ we obtain singlets (in which case the method is
  identical to CEP) and, generally, the operation succeeds with a
  finite probability given by Eq.~(\ref{eq:PCM-P}). For non-identical
  PMSs PCM yields two non-maximally entangled pure states which are
  then used for entanglement swapping followed by the Procrustean
  method. The probability of succeeding in converting both of the
  bonds to pure states is
\begin{equation}
p_c^2 = \lambda^2 \nu^2 (\alpha (1-\beta)+\beta (1-\alpha))^2.
\label{Equ:MSwap-EPerc-SPProb}
\end{equation}
These pure states have largest Schmidt coefficient
\begin{equation}
\hat \alpha = \frac{\max (\alpha (1-\beta),\beta (1-\alpha))}{(\alpha (1-\beta)+\beta (1-\alpha))}.
\label{Equ:MSwap-EPerc-State}
\end{equation}
So, by using the SCP in single edged swapping with pure states we find
that we can convert this pair of states into a singlet between the
end nodes with probability
\begin{equation}
2(1-\hat \alpha) = 2\frac{\min (\alpha (1-\beta),\beta (1-\alpha))}{(\alpha (1-\beta)+\beta (1-\alpha))}.
\label{Equ:MSwap-EPerc-SProb}
\end{equation}
Hence, the overall probability of succeeding with this scheme is
 \begin{align}
p_h = &2\lambda^2 \nu^2 [\alpha (1-\beta)+\beta (1-\alpha)]\nonumber \\
 &\times \min [\alpha (1-\beta),\beta (1-\alpha)].
\label{Equ:MSwap-EPerc-Prob}
\end{align}
\end{description}

If we compare the success probability of direct swapping, $p_d$, to
the probability of success in the classical percolation scheme,
$p_{CEP}$, it can be seen that classical percolation is more likely to
succeed in producing a singlet if
\begin{equation}
2\min (\alpha (1-\beta),\beta (1-\alpha))>\max (\alpha (1-\beta),\beta (1-\alpha)).
\label{Equ:MSwap-ImpInequality}
\end{equation}
But the ratio of the success probability for the classical scheme
against the hybrid protocol, $p_h$, is
\begin{equation}
\frac{p_{CEP}}{p_h} = \frac{2\min (\alpha (1-\beta),\beta (1-\alpha))}{(\alpha (1-\beta)+\beta (1-\alpha))}.
\label{Equ:MSwap-HPImpInequality}
\end{equation}
Whenever $\alpha\neq\beta$ this is less than one and there is an
improvement over the classical percolation scheme. Furthermore, the
hybrid scheme is more likely to succeed than direct swapping.
In Fig.~\ref{fig13} we compare the probabilities of success for all
schemes. As can be seen, hybrid swapping leads to the highest success
probability.
\begin{figure}[t]
  \centering\includegraphics[width=6cm]{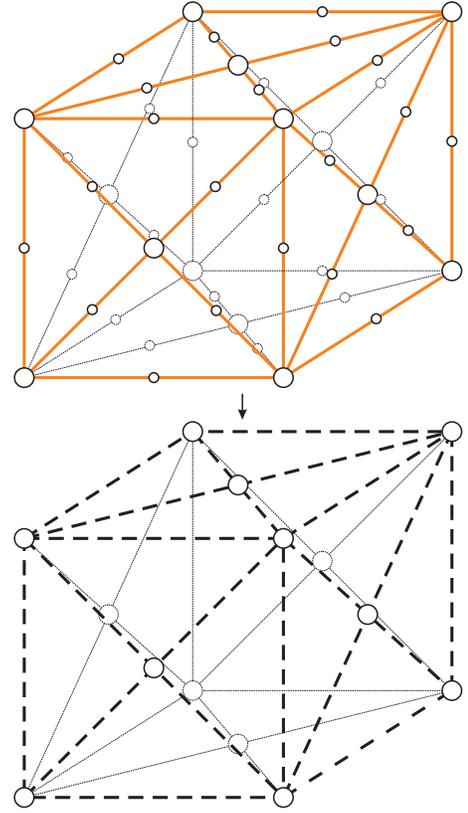}
\caption{%
  Illustration of entanglement percolation in a 3D network. The
  circles represent nodes containing qubits and the lines represent
  bonds containing pairs of two-qubit entangled states (the edges are
  not shown). The 3D network can be transformed into a Face-centered
  cubic network by performing the swapping operations (see
  Fig.~\ref{fig9}) over the smaller nodes. For some bond parameters
  the hybrid scheme allows percolation to occur where classical
  percolation fails.}
\label{fig14}
\end{figure}

Hybrid swapping can be used in sections of larger networks to
allow percolation to take place. A simple example is a face-centered
cubic (FCC) network, where every bond is split into two 2-edged bonds
(see Fig.~\ref{fig14}). When the above schemes are applied at the
nodes linking two 2-edged bonds the FCC network is recovered. Percolation is possible in
these 3D networks with a threshold of approximately $\approx 0.12$.
Since the classical scheme always gives a smaller success probability
than the hybrid scheme there are cases where the hybrid scheme
allows the percolation threshold to be exceeded but the classical
scheme does not (see Fig.~\ref{fig13}).

\subsection{Square Protocol}

\begin{figure}[t]
  \centering\includegraphics[width=6cm]{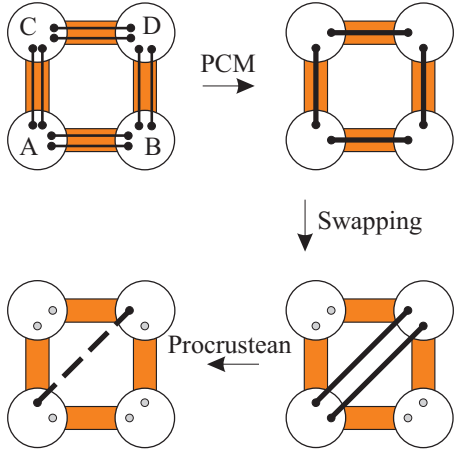}
\caption{%
  Application of the hybrid scheme in a square network. This involves
  transforming the PMSs into pure states probabilistically and then
  applying a suitable pure state procedure (see text). In the case
  shown all of the conversions are successful. When this happens a
  swapping operation can be performed and the resulting states
  distilled into a singlet.}
\label{fig15}
\end{figure}

\begin{figure}[t]
  \centering\includegraphics[width=8cm]{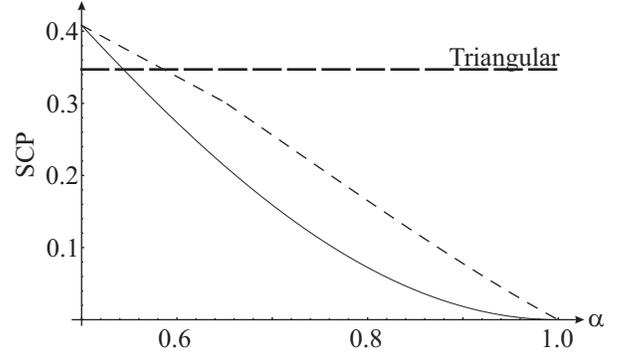}
\caption{%
  Comparison of singlet conversion probabilities for the different
  strategies in the square configuration, i.e. $\tilde p_{CEP}$ (solid
  line) and $p_{sq}$ (dashed line) for $\lambda=\nu=0.98,\,\beta=0.5$.  We
  have also indicated the percolation threshold for a triangular
  network. }
\label{fig16}
\end{figure}

CEP can also be improved on by using the hybrid strategy in a 2D
square network, as shown in Fig.~\ref{fig15}. Each bond is converted
into a pure state, $\ket{\hat\alpha}\equiv \sqrt{\hat\alpha} \ket{00}
+ \sqrt{1-\hat\alpha}\ket{11}$, by using PCM which is successful with
a probability $p_c$ on each bond. If this yields only two states
$\ket{\hat\alpha}$ having a common node ($B$ or $C$), entanglement
swapping can be performed followed by the procrustean scheme. If all
four PCMs succeed the resulting states can be connected (e.g. at nodes
$B$ and $C$) via a slightly modified version of entanglement swapping,
the so-called \emph{XZ-swapping}~\cite{Perseguers08}. For this
swapping operation the Bell measurement that usually has both qubits
measured in the $Z$ basis now measures one in the $X$ basis. After
this measurement unitaries are again applied to return the state into
Schmidt form. The results of the Bell measurement have an equal
probability, $p_m = 1/4$, for all outcomes $m$. Performing this
operation twice on the square leads to two pure states (between $A$
and $D$) of the form $\ket{\tilde\alpha}$, with $\tilde\alpha=
(1+\sqrt{1-16\hat\alpha^2(1-\hat\alpha)^2})/2$. These can be distilled
into a singlet with probability $\min[1,2(1-\tilde{\alpha}^2)]$ by
using the protocol based on Majorization~\cite{Nielsen01}. The overall
chance of succeeding in generating a singlet is then given by
\begin{equation}
p_{sq} = 4p_c^2(1-p_c^2)(1-\hat{\alpha}) + p_c^4\min (1, 2(1-\tilde{\alpha}^2)).
\label{Equ:SquareProb}
\end{equation}
When attempting to accomplish the same scheme using CEP we succeed
with a probability of $\tilde p_{CEP} = 1-(1-p_{CEP})^2$ which can be
significantly smaller than Eq.~(\ref{Equ:SquareProb}), as shown in
Fig.~\ref{fig16}.

\begin{figure}[t]
  \centering\includegraphics[width=8cm]{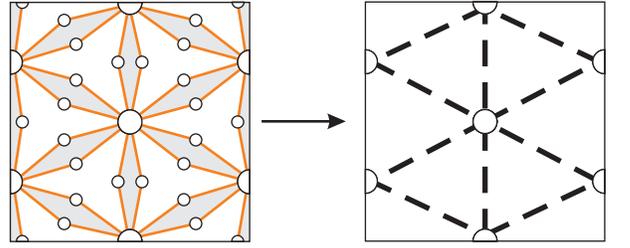}
\caption{%
  By applying the square protocol on the shaded regions a triangular
  network of randomly distributed singlets (dashed lines in the right
  figure) is recovered. In the left figure the nodes (circles) are
  linked by bonds (solid lines) each containing two edges (not
  shown).}
\label{fig17}
\end{figure}

Again, this improved strategy may enable an infinite cluster to form
when applied to larger networks. An example is shown in
Fig.~\ref{fig17}. Here the square protocol recovers a triangular
lattice. If the conversion of the squares into singlets succeeds with
a probability exceeding the percolation threshold an infinite cluster
forms. In Fig.~\ref{fig16} it can be seen that the hybrid scheme
exceeds the threshold for a triangular lattice in cases where CEP does
not.

\subsection{Hierarchical Networks}

Small networks like the square configuration discussed above can be
extended to larger networks in an iterative fashion. Networks formed
in this way from pure states were considered in~\cite{Perseguers08}.
Again the probability of successfully creating a singlet was shown to
be larger when quantum strategies were used instead of CEP. However,
the scheme with the highest probability is still unknown for these
`hierarchical' networks. Here we will consider two different
hierarchical networks with two edges per bond. Each of these contains
the square network at some iteration level. We determine the
SCP when using CEP in both cases which we then compare to the
hybrid strategy. As it turns out, the hybrid scheme outperforms CEP.

\begin{figure}[t]
  \centering\includegraphics[width=6cm]{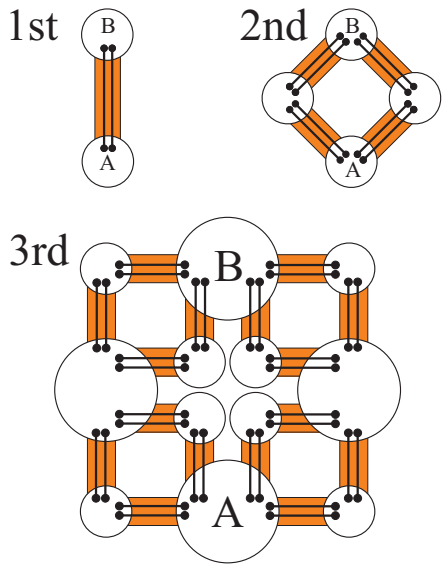}
\caption{%
  The first three iterations of a diamond lattice. We aim to create a singlet
  between nodes $A$ and $B$ in each case.}
\label{fig18}
\end{figure}

The first hierarchical network we consider is based on the `Diamond'
lattice, which at each stage replaces its bonds by the square network.
The geometry for the first three iterations is shown in
Fig.~\ref{fig18}.
The aim is to create a singlet between $A$ and $B$ and if we apply CEP
the probability of succeeding at each level is given by the iterative
formula
\begin{equation}
 p_{i}^{Diamond} = 1 - (1 - p_{i-1}^2)^2,
\label{Equ:DiamondIt}
\end{equation}
starting with $p_1^{Diamond} = p_{conv}$.

\begin{figure}[t]
  \centering\includegraphics[width=6cm]{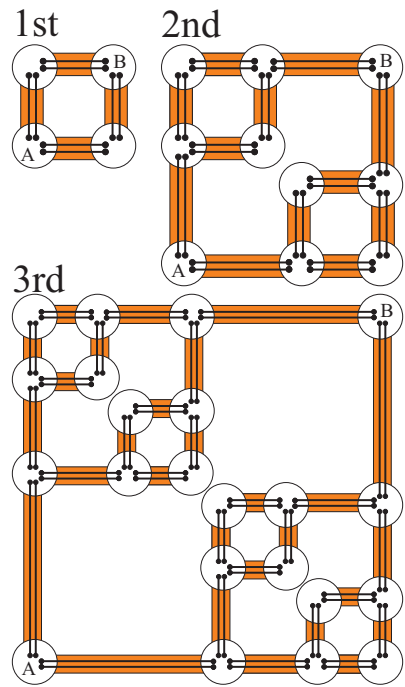}
\caption{%
  First three iterations of the tree lattice.  Each iteration is given
  by repeating the previous lattice twice and linking the pair of
  previous endpoints at new endpoints. We aim to create a singlet
  between nodes $A$ and $B$.  }
\label{fig19}
\end{figure}

The second hierarchical network we consider is the `Tree' network
which is again built on the square configuration. For these networks
an iteration is formed by creating two copies of the previous
iteration and linking the bottom-left and top-right corner of the
square to separate nodes $A$ and $B$ as shown in Fig.~\ref{fig19}.
Again, we wish to generate a singlet between the opposite corner nodes
($A$ and $B$) and CEP generates a singlet with a probability
\begin{equation}
 p_{i}^{Tree} = 1 - (1 - p_{i-1} p_{conv}^2)^2,
\label{Equ:TreeIt}
\end{equation}
where $p_0 = 1$.

Now we wish to see whether the hybrid scheme gives a larger SCP in
these networks. Once again, the hybrid scheme we consider starts
by converting all of the bonds into identical non-maximally entangled
pure states probabilistically. If the conversion fails on a bond then
the bond is destroyed. This results in a network containing random
pure state bonds. Each of these bonds contains one edge. Ideally we
would then apply a pure state protocol yielding the highest SCP
between the intended nodes, however, this protocol is not known in the
general case~\cite{Perseguers08}. Instead we apply a procedure which
performs $XZ$-swapping in cases when two bonds each have a qubit in
the same node (except if these nodes are $A$ or $B$). However, we also
distill pure states into states with more entanglement whenever two
edges form between two nodes and before performing further swapping.
Finally, once one state is obtained between $A$ and $B$, the
procrustean procedure is used to create a singlet.

\begin{figure}[t]
  \centering\includegraphics[width=8cm]{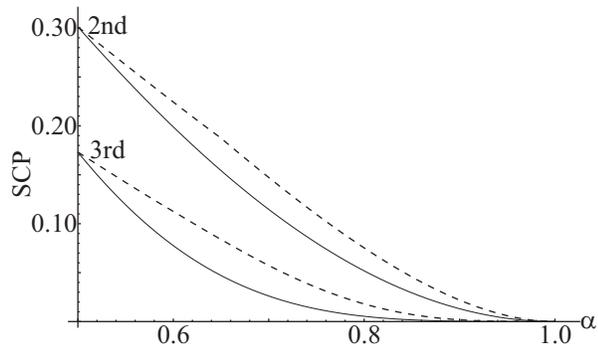}
\caption{%
  Probability of succeeding in generating a singlet between the
  endpoints of a diamond lattice for the 2nd and 3rd iterations
  (dashed lines). These give higher probabilities than the classical
  protocol (solid lines). The bonds contain two edges with parameters
  $\lambda=\nu=0.9,\,\beta=0.5$.  }
\label{fig20}
\end{figure}

\begin{figure}[t]
  \centering\includegraphics[width=8cm]{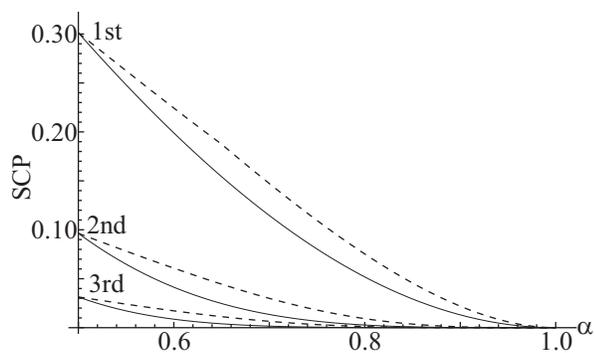}
\caption{%
  Probability of succeeding in generating a singlet between the
  endpoints of the 1st, 2nd and 3rd iterations of the tree lattice
  (dashed lines). These also outperform the classical protocol (solid
  lines). The bonds contain two edges with parameters
  $\lambda=\nu=0.9,\,\beta=0.5$.  }
\label{fig21}
\end{figure}

We applied this protocol to the hierarchical diamond and tree
networks. For the second and third iterations of the diamond lattice
the probabilities of creating a singlet are given in Fig.~\ref{fig20}
together with the probabilities using CEP. This comparison was also
made for the first, second and third iteration of the tree network and
the results are shown in Fig.~\ref{fig21}. These examples all
illustrate an improvement in the probability of forming a singlet when
using the hybrid method rather than classical percolation.

\section{Conclusion}\label{sec6}
We have demonstrated that within lattice networks, where the nodes are
connected by multiple bipartite mixed states, percolation strategies
can be applied for distributing entanglement. This is reliant on the
states being PMSs, which are known from the DLCZ repeater scheme and
arise as a result of amplitude damping. To show this we have introduced
some new purification protocols designed to maintain the form of these
states or generate singlets. Like in the pure state case, a higher
probability of distributing a singlet can be obtained, when the states
in a bond are not identical. The question of whether quantum
strategies can outperform CEP when each edge in a bond is identical is
still open. Since we have shown that classical entanglement percolation is only
possible for a specific class of bipartite states, entanglement distribution in
a network which is subject to more general forms of noise needs to
make use of other methods. The development of these methods is one of
the most important goals for future work.  These will not produce
perfectly entangled states, however, the resulting state fidelity may
be independent of distance and sufficient for purification. An example
of such a strategy is given in Ref.~\cite{Perseguers08b} for a bit-flip
noise model. Progress in this direction has also been accomplished by
generating 3D thermal cluster states using Werner
states~\cite{raussendorf05,perseguers09c}.

\acknowledgments
This research was supported by the EPSRC (UK) through the QIP IRC
(GR/S82176/01) and the ESF project EuroQUAM (EPSRC grant
EP/E041612/1).

\appendix

\section{Proof of singlet distillation requirement \label{app1}}

In this appendix we give a concise proof for a necessary and
sufficient condition to be able to create a singlet out of entangled
mixed states using LOCC. We allow the states to be arbitrary bipartite
states which are shared between two nodes and all operations are LOCC.
A similar proof, but partly restricted to identical states, was given in~\cite{Jane02}.

\begin{lem}\label{lem1}
  If a quantum state $\rho_{ab}$ can be distilled to a pure states, $\ket{\Psi}$, then any state with
  the same range $R(\rho_{ab})$ is also distillable to this state with non-zero probability.
\end{lem}

\begin{proof}\label{proof1}
  The general form of the state is
\begin{equation}\rho_{ab} = \sum_{i=1}^N p_i \ket{\psi_i}\bra{\psi_i},\end{equation}
with $ p_i>0$, $\sum_i p_i = 1$ and $\ket{\psi_i}\in H_A \otimes H_B $.
If the state is distillable to a pure state, $\ket{\Psi}$, there exist
linear operators $M_A$ and $N_B$, with $M_A M_A^\dagger \leq I, N_B
N_B^\dagger \leq I $, such that
\begin{align}
&M_A \otimes N_B \rho_{ab} M_A^\dagger \otimes N_B^\dagger = p \ket{\Psi}\bra{\Psi} \\
\Rightarrow  &M_A \otimes N_B \ket{\psi_i}\bra{\psi_i} M_A^\dagger \otimes N_B^\dagger \varpropto \ket{\Psi}\bra{\Psi}
\end{align}
or
\begin{equation}M_A \otimes N_B \ket{\psi_i}\bra{\psi_i} M_A^\dagger \otimes N_B^\dagger = 0.\end{equation}
This can be summarized as
\begin{equation}\Rightarrow M_A \otimes N_B \ket{\psi_i} = q_i\ket{\Psi},\end{equation}
where at least one $q_i$ is non-zero as otherwise the operator fails to distill $\rho_{ab}$. If this condition is satisfied the operation distills the mixed state into $\ket{\Psi}$. Now given another state $\tilde{\rho_{ab}}$ with the same range as
$\rho_{ab}$. We have that
\begin{equation}
\tilde{\rho_{ab}} = \sum_{i=1}^M \tilde{p_i} \ket{\tilde{\psi_i}}\bra{\tilde{\psi_i}}
\end{equation}
with $ \ket{\tilde{\psi_i}} = \sum_{j=0}^N a_{i,j} \ket{\psi_j}$ and $ \ket{\psi_i} = \sum_{j=0}^M b_{i,j} \ket{\tilde{\psi_i}}$. This then gives
\begin{align}
M_A \otimes N_B \ket{\tilde{\psi_i}} &=  M_A \otimes N_B  \sum_{j=0}^N a_{i,j} \ket{\psi_j} \\
&= \sum_{j=0}^N a_{i,j}q_j \ket{\Psi} = \tilde{q_i}\ket{\Psi}.
\end{align}
The value of one $\tilde{q_i}$ must be non-zero as otherwise all
$q_i$ are zero and this contradicts the fact that the operator distills $\rho_{ab}$. Hence the protocol
also distills $\tilde{\rho_{ab}}$.
\end{proof}

\begin{lem}\label{lem3}
  For $n$ 2-qubit states to be distillable into a pure singlet at
  least two 2-qubit states cannot have a range spanned by product states.
\end{lem}

\begin{proof}\label{proof3}
  If a 2-qubit state has a range that can be spanned by product states
  then a separable state with this range exists. If there are $n$
  states each with a range spanned by product states the system state
  would have a range equivalent to a separable state formed by all of
  these 2-qubit separable states. Since it is impossible to distill a
  pure entangled state from any separable state it is impossible to
  distill a pure entangled state from $n$ two qubit states each with a
  range spanned by product states. Similarly, if one of the 2-qubit
  states does not have a range spanned by product states, but all of
  the other states do, the range is equivalent to the range formed
  from a separable state and one mixed 2-qubit state. This can not be
  distilled into a pure singlet as it would contradict the result in
  \cite{Kent98}.  Hence at least two states can not have a range
  spanned by product states to be able to distill a pure entangled
  state.
\end{proof}

We now need to look at the two qubit states that satisfy this
property. The states with rank one are already pure and if they have
rank four the range can be spanned by product states. Similarly, if the state
has rank three it can also be spanned by product states. This can be
seen by considering the subspace orthogonal to a general state
$\sqrt{\alpha}\ket{00}+\sqrt{1-\alpha}\ket{11}$. This space is spanned
by $\{ \ket{01} ,\, \ket{10} ,\,
(\sqrt{1-\alpha}\ket{0}-\sqrt{\alpha}\ket{1})(\ket{0}+\ket{1]})/\sqrt{2}\}$
and these are all product states. The last states to consider are those of rank 2, which fall into two
categories \cite{Sanpera98}. The range is either spanned by product states
$\{ \ket{00}
,\,(\sqrt{\lambda}\ket{0}-\sqrt{1-\lambda}\ket{1})(\sqrt{\mu}\ket{0}+\sqrt{1-\mu}\ket{1})\}$
or $\{ \ket{00}
,\,(\sqrt{\alpha}\ket{01}+\sqrt{\beta}\ket{10})+\sqrt{1-\beta-\alpha}\ket{00})\}$.
Hence only states that have a range containing one product state are
the mixed states satisfying the condition. All mixed rank two states of two qubits can be considered to be the mixed state formed by tracing out a third qubit from
a pure three qubit system. The classifications of these 3 qubit
systems is given in \cite{Acin00,Dur00,Acin01} and for the range of
the mixed system to contain one product state the three qubit state
belongs to the W class. This class can always be written as
$\sqrt{\lambda} \ket{\Phi}\ket{1} + \sqrt{1-\lambda}
\ket{00}\ket{0} $ with
\begin{equation}
\ket{\Phi}=\sqrt{\alpha}\ket{01}+\sqrt{\beta}\ket{10}+\sqrt{\gamma}\ket{00},\,\alpha+\beta+\gamma=1.
\end{equation}
By tracing out one qubit and using local operations the 2-qubit state
that can not be spanned by product states has the form
\begin{equation}
\rho=\lambda\ket{\psi}\bra{\psi}+(1-\lambda)\ket{01}\bra{01}),
\end{equation}
where
\begin{equation}
\ket{\psi}=\sqrt{\alpha}\ket{00}+\sqrt{\beta}\ket{11}+\sqrt{\gamma}\ket{01}, \, \alpha+\beta+\gamma=1.
\end{equation}
So the only states that can be purified into a perfect singlet, given
finite copies, are of this form.

If there are two states of this form we know that the system is
distillable since the procedure given in Sec.~\ref{sec2} succeeds in
the distillation.

\section{The Distillable Subspace Scheme \label{app2}}
To extend the DSS scheme to $n$ PMSs, $\rho(\alpha,\lambda)$, we first
need to describe the $2^n$ non-zero eigenvalues and their
eigenvectors. These correspond to different combinations of $n-l$
$\ket{\alpha}$ terms and $l$ $\ket{01}$ terms. Then taking the decimal
representation of the local states we can label each of these
eigenvectors by the decimal difference between the values at each
location. This difference $y$ in binary gives the location of the
$\ket{01}$ terms. For example, in the case of two identical PMSs these
are

\begin{eqnarray}
y=0=00:& \ket{\alpha}\ket{\alpha},  \nonumber \\
y=2=10:& \ket{01}\ket{\alpha}, \nonumber \\
y=1=01 :& \ket{\alpha}\ket{01}, \nonumber \\
y=3=11 :& \ket{01}\ket{01}  \label{Equ:GPEigenSysK}
\end{eqnarray}
and $y$ takes all of the values from 0 to $2^n - 1$. Now we define
$m(x)$ to be the number of 1s in the binary representation of $x$ and
$T_y = \lbrace x $ : $ x ^{\wedge} y = 0,\, 0\leq x< 2^n,\, x \in
\mathbb{N} \rbrace $ (`$^{\wedge}$' is the bitwise AND
  operation).

Then $l=m(y)$ and all of the terms in a non-zero eigenvector are of
the form
\begin{equation}
\sum_{x\in T_y} \sqrt{\alpha^{n-m(x)-l} (1-\alpha)^{m(x)}} \ket{x}_{A}^d \ket{x+y}_{B}^d,
\label{Equ:GP-EvDesc}
\end{equation}
with eigenvalue $ \lambda^{n-l} (1-\lambda)^{l}$.

From this structure we can project out an entangled state if we
measure the operator $(\ket{c}_A^d \bra{c}_A^d + \ket{d}_A^d
\bra{d}_A^d )$ at $A$ and then $(\ket{c+y}_B^d \bra{c+y}_B^d +
\ket{d+y}_B^d \bra{d+y}_B^d )$ at $B$, when $ c\in T_y$, $d \in T_y $, $d>c$ and as long
as there are no other terms of the form $\ket{c}_A^d\ket{d+y}_B^d +
\ket{d}_A^d\ket{c+y}_B^d$, $\ket{c}_A^d\ket{d+y}_B^d$ or
$\ket{d}_A^d\ket{c+y}_B^d$ in any non-zero eigenstate.

The term $\ket{c}_A^d\ket{d+y}_B^d + \ket{d}_A^d\ket{c+y}_B^d$ can not appear
in one eigenstate since all of the terms must have the same $y$ value
and this would require $c$ to be equal to $d$.
The state $\ket{c}_A^d\ket{d+y}_B^d$ lies in one if and only if $\exists w\in
\mathbb{N}$, $0\leq w\leq 2^n - 1$ such that $c+w=d+y$ and $c\in T_w$. Similarly for $\ket{d}_A^d\ket{c+y}_B^d$ but this  case can not occur since $ c\in T_y$ and $d \in T_y $ means that $w\geq y$ and $d+w>c+y$. If we assume that
$\exists w\in \mathbb{N}$, $0\leq w\leq 2^n - 1$ such that $c+w=d+y$
and $c\in T_w$ this would mean that $d=c+k$ and that $c^\wedge k=0$ for some $k>0$.
Both of these results then give that $k^\wedge y=0$ and $w=k+y$. So, if $w=y+k=d+y-c$ such that $c^\wedge k=0$ can not
be satisfied we create a maximally entangled state.

Now we have a choice of ways of creating these measurements. One
particular way involves the definition of sets $S_k = \lbrace x$ :
$m(x)=k$, $ 0\leq x< 2^n,\,x\in \mathbb{N} \rbrace$ and $J_{a,b}=\lbrace
x $: $x ^{\wedge}$($a$ OR $b$) = 0, $ 0\leq x< 2^n, x\in \mathbb{N}
\rbrace$. Then the protocol consists of performing a POVM
$P_{k,a,b}=C_k(\ket{a}_A^d\bra{a}_A^d + \ket{b}_A^d\bra{b}_A^d)$ at
location $A$ with $a,\,b\in S_k$, $a\neq b$ and $0<k<n$. For $k=0$ and $n$ we define
$P_{k,a,b}=\ket{2^k-1}_A^d\bra{2^k-1}_A^d$ and when these outcomes occur the procedure has failed. Here $C_k$ is a factor to
ensure that
\begin{equation}
\sum_{a,b\in S_k, k=0}^{k=n} P_{k,a,b}= I.
\label{Equ:GP-NormCond}
\end{equation}
With this outcome at location $A$ another POVM is done at location $B$
given by the operators $Q_d=C_k(\ket{a+d}_B^d\bra{a+d}_B^d +
\ket{b+d}_B^d\bra{b+d}_B^d)$ ($d \in J_{a,b}$) and $F=I-\sum_{d\in
  J_{a,b}} Q_{d}$. If the outcome here is $F$ the protocol has failed,
otherwise we have obtained a maximally entangled state. This protocol
works since $ a,\,b \in T_y $ for all $ y \in J_{a,b} $ but there is no
$w=k+y$ such that $k^\wedge a=0$ and $b+y=a+y+k$, since if there
were we would have $b=a+k$ but $m(a+k)\neq m(b)$.

The probability of succeeding is given by Eq.~({\ref{Equ:GP-ProbFn}})
which comes from considering a particular eigenstate with parameter
$y$. In this eigenstate there are
\begin{equation}
 N_1 = \binom{n-m(y)}{k}\left(\binom{n-m(y)}{k}-1\right)/2
 \label{Equ:GP-ProbFnWork1}
\end{equation}
different pairing terms, $\ket{a}_A^d\ket{a+y}_B^d +
\ket{b}_A^d\ket{b+y}_B^d$ with $a,b \in S_k$. The number of possible
measured operators from this eigenstate is given by
\begin{equation}
N_2 = \binom{n-m(y)}{k}\left(\binom{n}{k}-1\right).
 \label{Equ:GP-ProbFnWork2}
\end{equation}
Note that the pairings in the eigenstate are twice as likely to occur
than the ones with just an overlap and these have been counted twice
in this sum. The probability that starting with an eigenstate
(parameter $y$) we succeed is then
\begin{equation}
\frac{2N_1}{N_2} = \frac{\binom{n-m(y)}{k}-1}{\binom{n}{k}-1},
 \label{Equ:GP-ProbFnWork3}
\end{equation}
given that we have measured the operator $S_k$ and the probability of this
was
\begin{equation}
P_k = \binom{n-m(y)}{k}\alpha^{n-m(y)-k}(1-\alpha)^k.
 \label{Equ:GP-ProbFnWork4}
\end{equation}
By summing over these we have, given we start with an eigenstate
with $m(y)=m$, the probability of succeeding to be
\begin{equation}
\sum_{k=1}^{n-m-1} \frac{\alpha^{n-m-k} (1-\alpha)^k \binom{n-m}{k} \left(\binom{n-m}{k}-1\right)}{\binom{n}{k}-1}
\label{Equ:GP-ProbFnWork5}
\end{equation}
and these eigenstates occur with probability
\begin{equation}
\lambda^{n-m} (1-\lambda)^m \binom{n}{m}.
\label{Equ:GP-ProbFnWork6}
\end{equation}
We have not counted $k=0,\,n-m$ since they never contribute to the
success probability. Then by summing over all of these we get the result in
Eq.~(\ref{Equ:GP-ProbFn}).


\end{document}